\documentclass[pdflatex,sn-mathphys]{sn-jnl}

\usepackage{amsmath}
\usepackage{amssymb}
\usepackage{latexsym}
\usepackage{url}
\usepackage{xcolor}
\usepackage{booktabs}

\usepackage[binary-units = true, free-standing-units=true]{siunitx}
\DeclareSIUnit\px{px}

\usepackage[binary-units = true, free-standing-units=true]{siunitx}
\DeclareSIUnit\px{px}
\usepackage{mathtools}

\usepackage{hyperref}

\jyear{2023}%

\begin{document}

\title[Improved distinct bone segmentation using multi-resolution networks]{Improved distinct bone segmentation in upper-body CT through multi-resolution networks.}


\author*[1]{\fnm{Eva} \sur{Schnider}}\email{eva.schnider@unibas.ch}

\author[1]{\fnm{Julia} \sur{Wolleb}}
\author[1]{\fnm{Antal} \sur{Huck}}
\author[2]{\fnm{Mireille} \sur{Toranelli}}
\author[1]{\fnm{Georg} \sur{Rauter}}
\author[2]{\fnm{Magdalena} \sur{Müller-Gerbl}}
\author[1]{\fnm{Philippe C.} \sur{Cattin}}

\affil*[1]{\orgdiv{Department of Biomedical Engineering}, \orgname{University of Basel}, \orgaddress{\street{Hegenheimermattweg 167B}, \city{Allschwil}, \postcode{4123}, \country{Switzerland}}}

\affil[2]{\orgdiv{Department of Biomedicine, Musculoskeletal Research}, \orgname{University of Basel}, \orgaddress{\city{Basel}, \country{Switzerland}}}




\abstract{\textbf{Purpose:} Automated distinct bone segmentation from CT scans is widely used in planning and navigation workflows.
U-Net variants are known to provide excellent results in supervised semantic segmentation. However, in distinct bone segmentation from upper body CTs a large field of view and a computationally taxing 3D architecture are required. This leads to low-resolution results lacking detail or localisation errors due to missing spatial context when using high-resolution inputs.

\textbf{Methods:} We propose to solve this problem by using end-to-end trainable segmentation networks that combine several 3D U-Nets working at different resolutions. Our approach, which extends and generalizes HookNet and MRN, captures spatial information at a lower resolution and skips the encoded information to the target network, which operates on smaller high-resolution inputs. We evaluated our proposed architecture against single resolution networks and performed an ablation study on information concatenation and the number of context networks.

\textbf{Results:} Our proposed best network achieves a median DSC of 0.86 taken over all 125 segmented bone classes and reduces the confusion among similar-looking bones in different locations. These results outperform our previously published 3D U-Net baseline results on the task and distinct-bone segmentation results reported by other groups.

\textbf{Conclusion:} The presented multi-resolution 3D U-Nets address current shortcomings in bone segmentation from upper-body CT scans by allowing for capturing a larger field of view while avoiding the cubic growth of the input pixels and intermediate computations that quickly outgrow the computational capacities in 3D. The approach thus improves the accuracy and efficiency of distinct bone segmentation from upper-body CT.
}




\maketitle

\section{Introduction}\label{sec1}

Segmentation of bones is used in bone disease diagnosis,  in image-based assessment of fracture risks \cite{deng2022deep}, bone-density \cite{uemura2022development}, for planning and navigation of interventions \cite{su2022three}, and for post-treatment assessment. 

Bone tissue segmentation from CT has been shown to work well using slice-wise 2D CNN-based segmentation algorithms \cite{klein2019automatic,leydon2021bone,noguchi2020bone}.
The tasks and solutions become more varied when moving from bone-tissue segmentation to distinct bone segmentation (our task) where we distinguish individual bones. 
Vertebrae segmentation has gained much attention, with many of the algorithms using multi-stage approaches and leveraging the sequential structure of the spine \cite{payer2020coarse}. 
Rib segmentation has been tackled by \cite{yang2021ribseg}, who use a point cloud approach targeted at leveraging their dataset's spatial sparsity.
Carpal bone segmentation is performed from X-rays of hands that were placed on a flat surface \cite{faisal2021carpal}.

Simultaneous segmentation of distinct bones of multiple groups is still relatively little studied.
\cite{fu_hierarchical_2017} segment 62 different bones from upper-body CT using an atlas-based approach and kinematic joint models.
\cite{lindgren_belal_deep_2019} use a multi-stage approach with a localisation network, shape models, and a segmentation network to segment 49 distinct bones of the upper body.
Segmentation of bones of different groups in one shot can be used as a starting point for more fine-grained atlas segmentations \cite{fu_hierarchical_2017}, or as a guide for a follow-up inner organ segmentation \cite{kamiya2018automated}. Segmenting multiple structures at once can also be beneficial for the segmentation accuracy, \cite{boutillon2022multi} found their network trained on multiple bone classes to outperform the one-class networks.

The region of interest in upper-body or full-body CT scans is typically larger than the possible input sizes of 3D convolutional neural networks (CNNs).
As a result, the input needs to be sampled as patches, restricting the input field of view to the patch size.
This problem exacerbates with the development of CT scanners that produce ever more highly resolved images. While a higher resolution allows for capturing more fine-grained details, it covers smaller body areas within a fixed-size input patch.

In order to extend the field of view, larger input patches can be sampled.
Using bigger patches, i.e. more input pixels does not increase the number of trainable parameters in a fully connected network, but it does increase the number of necessary intermediate computations. 
Doubling the patch size in all three dimensions leads to at least eight times more forward- and backward computations, which are taxing for the generally scarce GPU memory. 
Countermeasures fall into two categories. A) keeping the resolution and input pixel size high, but reducing the computational load elsewhere. Those measures include reducing the batch size (not to be confused with the patch size), using a simpler model, or reducing the output size. 
All of those means potentially hamper training and inference.
B) Keeping a large field of view by using a small patch size of down-sampled inputs. This approach allows for a wider field of view for a constant input size while losing detail information.

\begin{figure*}[!t]
\centering
\includegraphics[width=1.0\textwidth]{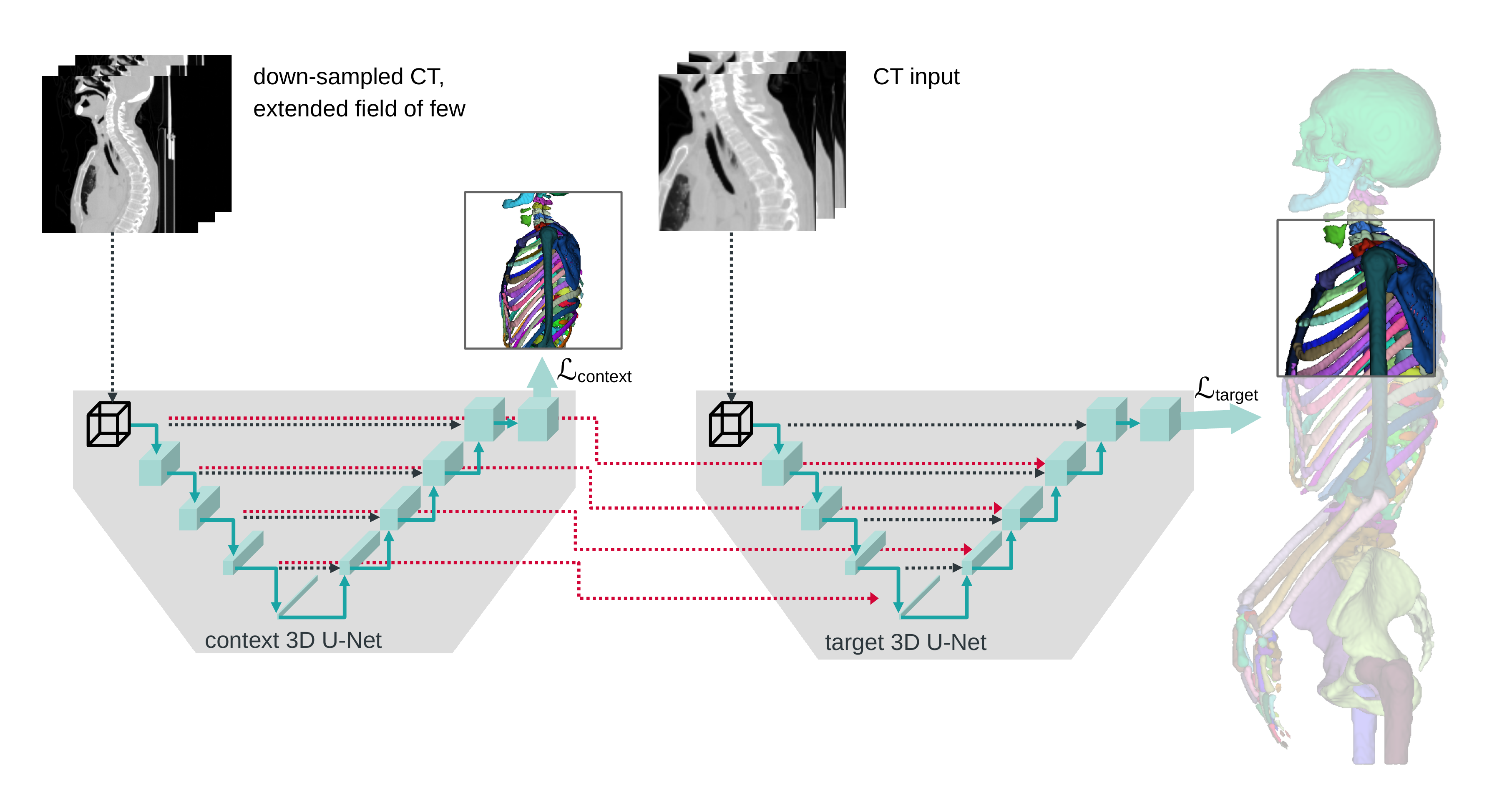}
\caption{Task overview: We segment 125 distinct bones from upper-body CT scans using SneakyNet, a multi-encoder-decoder network which incorporates inputs at various resolutions. The example here features one context network, but multiple are possible.}\label{fig_overview}
\end{figure*}

To decide upon the better of the two approaches presented above, the requirements for the task at hand need to be considered.
A suitable network for our task of complete distinct bone segmentation from upper-body CT scans (see \ref{fig_overview}) should provide the following: Its field of view should be sufficiently big to distinguish similar bones at different body locations, e.g. left from right humerus or the fourth from the eighth rib while keeping the computational burden in a feasible area.



The merits of high-resolution inputs -- accurate details -- and low-resolution inputs -- a larger field of view -- can be combined in many ways.
Cascaded U-Nets consist of two or more individual U-Nets that are trained consecutively. 
A first model is trained on downsampled input. Its one-hot encoded segmentation results are then upsampled, potentially cropped and used as additional input channels for the following model at higher resolution \cite{isensee2021nnu}.
These approaches all have the downside of requiring the training and sequential inference of multiple models. Instead of this, we focus on end-to-end trainable models here.

End-to-end trained multi-resolution architectures have been proposed in histopathology whole-slide segmentation. For example, MRN \cite{gu2018multi} combines a 2D target U-Net and one context encoder with drop-skip-connections crossing over at every level. MRN does not contain a context decoder or context loss and is studied on a binary segmentation problem. Another such architecture is HookNet \cite{van2021hooknet}, which contains both a target and a context 2D U-Net and two individual losses, but only uses skip connections in the bottleneck layer.

The purpose of our work is to address common segmentation errors that originate from a lack of global context while using 3D U-Nets for distinct bone segmentation. We propose to use a multi-resolution approach and present SneakyNet, an expansion and generalization of the MRN and HookNet architectures.
We compare the segmentation accuracy, complexity, and run-time of baseline 3D U-Nets with the SneakyNet. We ablate the model components and find that the use of our generalized architecture improves the results over the HookNet and MRN variants.
We will use our bone segmentation in conjunction with 3D rendering of anatomical images in augmented- and virtual reality applications, where segmentations can be used on top or in conjunction with existing transfer functions \cite{faludi2021transfer,zelechowski2021patient}.


\section{Materials and Methods}


To assess the performance of SneakyNet on upper-body distinct bone segmentation, we train it on our in-house upper-body CT dataset, which has been described in \cite{schnider2022improved}. We make ablation studies on the combination of context and target information and on the optimal number of context networks.


\begin{table}[!t]
\caption{\label{tab_parameter_comparison}Comparison of architectures with different field of view (FOV) of their target and context network(s).}
\centering
\begin{tabular}{lrlrrr}
\toprule
Config & Target  & Context  & trainable  & input  & training time \\
 &  network &  network(s) &  param. &  pixels &per iteration\\
  &  FOV &  FOV &  $\cdot10^7$  &  $\cdot10^4$  & (s) \\
\midrule

 A 3D U-Net & $32^3$ & --- & $5.8 $ &$3.3 $& 0.44\\
 & $64^3$ & --- &  &$26.2 $& 0.57\\
\phantom{A} 3D U-Net slim$^*$ & $128^3$ & --- & $1.5 $ &$209.7 $&~  4.24\\
B HookNet & $32^3$ & $64^3$ & $3.7 $ &$6.6 $& 0.41\\
 & $64^3$ & $128^3$ &  &$52.4 $& 0.72\\
C MRN & $32^3$ & $64^3$ & $4.7 $ &$6.6 $& 0.43\\
 & $64^3$ & $128^3$ &  &$52.4 $& 1.27\\
D Sneakynet (ours) & $32^3$ & $64^3$ & $4.9 $ &$6.6 $& 0.45\\
&   & $64^3-128^3$ &$5.8 $ &$9.9 $&0.70\\
  &   & $64^3-128^3-256^3$ & $6.2 $&$13.1 $& 3.16\\
  & $64^3$ & $128^3$ & $4.9 $ &$52.4 $& 1.28\\
  &   & $128^3-256^3$ &$5.8 $&$78.6 $& 3.11\\

\bottomrule
\multicolumn{6}{@{}l}{$^*$ Operating the full 3D U-Net on patches of size $128^3$ exceeds the available GPU memory.}
\end{tabular}
\end{table}

\subsection{SneakyNet Architecture} \label{sec_networks}

In general, SneakyNet consists of one target network and one or more context networks. 
The target network operates on high-resolution data and eventually produces the desired segmentation maps. The context networks operate on lower resolution inputs spanning a larger field of view. Information is propagated from the context networks to the target network using crop-skip connections presented in \autoref{sec_cropskip}. We present a detailed visual overview of the architecture with one context network in \autoref{fig_overview}.

\begin{figure*}[!t]
\centering
\includegraphics[width=\textwidth]{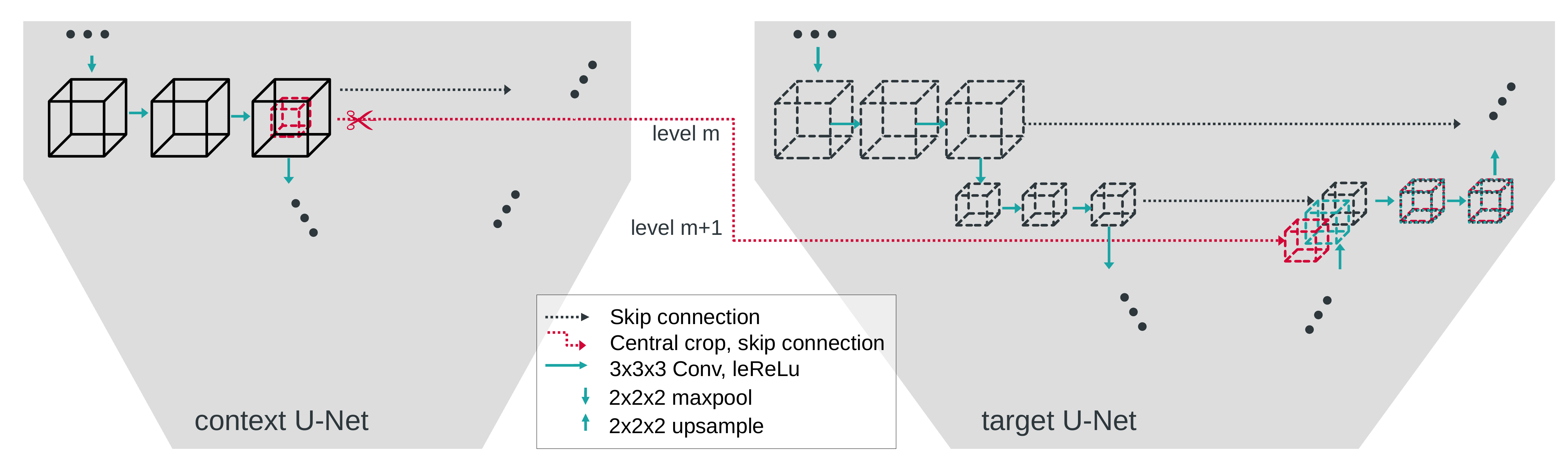}
\caption{Detailed view of the architecture. Displayed are only two out of five levels of the U-Nets. Left: the context U-Net working on low-resolution data with a larger field of view. Right: The U-Net working with the central cropped high-resolution data. After all encoder convolutions of level $m$, a cropped copy of the output is skipped to the target decoder at level $m+1$. The decoder receives skip connections from its own encoder and the context network. The intermediate results of the decoder and both skip connections are concatenated along the channel axis before undergoing further convolutions. }
\label{fig_crop_unet_detail}
\end{figure*}

In our previous work \cite{schnider20203d}, we have explored the suitability of different 2D and 3D network architectures and parameter configurations for upper-body distinct bone segmentation. We found that there is little leeway in architectural choices due to the tasks large required field of view and the many classes that are segmented in parallel. A lean 3D U-Net variant was found to work best \cite{schnider20203d}. We use this variant's architecture for our target and context U-Nets here. In our baseline computations, where we have only a target network and omit the context networks, we select the number of channels in order for our variants and the baselines to have approximately the same number of trainable parameters, to facilitate comparison.  
Inputs to the network are required to be multiples of $2^{M-1}$, where $M$ denotes the number of levels of the U-Net. We use the basic architecture of $M=5$ and therefore need multiples of 16 pixels in every dimension as input.

For the target network we use inputs of size $(Sx,Sy,Sz)$ at full resolution. For each of the context networks we use that input plus its surrounding area, which together span a field of view of $2^\kappa\cdot(Sx,Sy,Sz)$. We display the case of $\kappa=1$ in \autoref{fig_overview}, but also use context networks with $\kappa=2$ and $\kappa=3$ in our ablation studies.
The context network inputs are down-sampled to reduce their size to $(Sx,Sy,Sz)$. We perform the down-sampling using $(2^\kappa \times 2^\kappa \times 2^\kappa)$ average-pooling with a stride of $2^\kappa$. 
Both target and context network inputs eventually have a size of $(Sx,Sy,Sz)$, but at different resolutions and fields of view.

\begin{figure}[!t]
\centering
\includegraphics[width=1.0\textwidth]{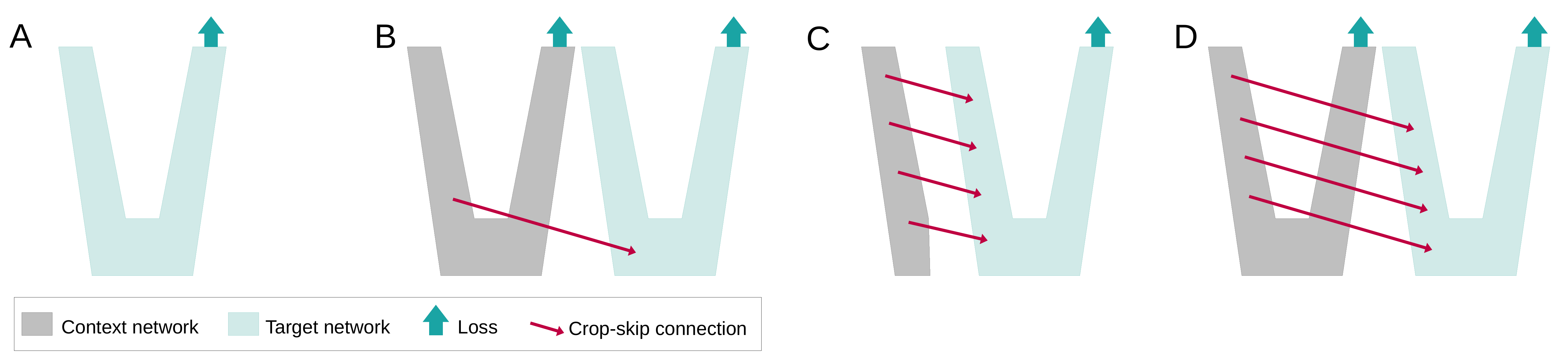}
\caption{Schematic of the four network configurations used in our ablation study. A shows a base U-Net, while B, C, D show different possibilities of how to insert information into the target network, see also \autoref{sec_cropskip} for a written description.}
\label{fig_ablation_networks}
\end{figure}

\subsubsection{Crop-skip connections}\label{sec_cropskip}
We use crop-skip connections to transfer information from the context to the target branch. We crop the encoder output at the desired level $m$ such that only the centre cube of half the size per dimension remains. This centre cube is now spatially aligned to the input of the target branch. We concatenate the centre cube to the next lower level $m+1$ of the target decoder to match the spatial size. We refer to the central cropping and subsequent concatenation into a lower level of the target branch as crop-skip-connection. A detailed schematic of the crop-skip connection is depicted in \autoref{fig_crop_unet_detail}.

We explore three network configurations which differ in their number of crop-skip connections and their use of a context loss, and compare it to a baseline U-Net. A visual comparison of the architectures is given in \autoref{fig_ablation_networks} and the parameters are provided in \autoref{tab_parameter_comparison}.

\begin{itemize} 
\item \textbf{A -- Baseline:} 
3D U-Net with optimal configuration found for the task \cite{schnider20203d}.
\item \textbf{B -- HookNet:} 
One context network with a single crop-skip connection is added to the target network. The crop-skip connection enters the target network at its bottleneck layer. This configuration is used in \cite{van2021hooknet}.
\item \textbf{C -- MRN:} 
Crop-skip connections connect the context encoder and the target decoder at every level. There is neither a context decoder nor a context loss function. This configuration was used in \cite{gu2018multi}.
\item \textbf{D -- proposed SneakyNet:} 
Crop-skip connections connect all levels of the context and target networks. The context network has a decoder with its own loss function.
\end{itemize}

\subsection{Training}
Our dataset is split into 11 scans for training, 2 for validation and 3 for testing. We use 5-fold cross-validation, ensuring that every scan appears in precisely one of the cross-validation folds in the test set.

The loss is composed of an unweighted combination of the target network's loss and the losses of the $K$ context networks.
For both networks, we use the sum of the cross-entropy loss $\mathcal{L}_{\text{X-Ent}}$ and Dice-Loss $\mathcal{L}_{\text{DSC}}$ \cite{milletari2016v}. As in \cite{schnider20203d}, we sum the Dice-Loss for every class separately and normalize by the number of classes. We optimized the network weights using the Adam optimizer with an initial learning rate of 0.001.
We trained our networks for 100000 iterations until convergence was observed.

Our input images are padded by $(S-S_{\mathrm{target}})/2$ all-around using edge value padding. The padding step ensures that we can sample high-resolution patch centres right to the image's border.

We implemented and trained our networks using Tensorflow Keras 2.5. All training and inference were conducted on NVidia Quadro RTX 6000 GPUs of \SI{24}{\giga \byte} RAM size.


\subsection{Evaluation}

We evaluate the performance of our models using a class-wise Dice Score Coefficient (DSC).
To indicate the performance over all classes, we give the median and the 16 and 84 quantiles ($1\sigma$) over all classes~$c$. To not give a distorted impression of the distribution, we exclude classes where no true positives of $c$ have been detected and therefore $\mathrm{DSC_c}=0$. We present the percentage of classes included as 'non-zero DSC' in \autoref{tab_ablation} and \autoref{tab_context_branch_comparison} to make up for the omission.

\section{Results and Discussion}

\begin{figure*}[!t]
\centering
\includegraphics[width=\textwidth]{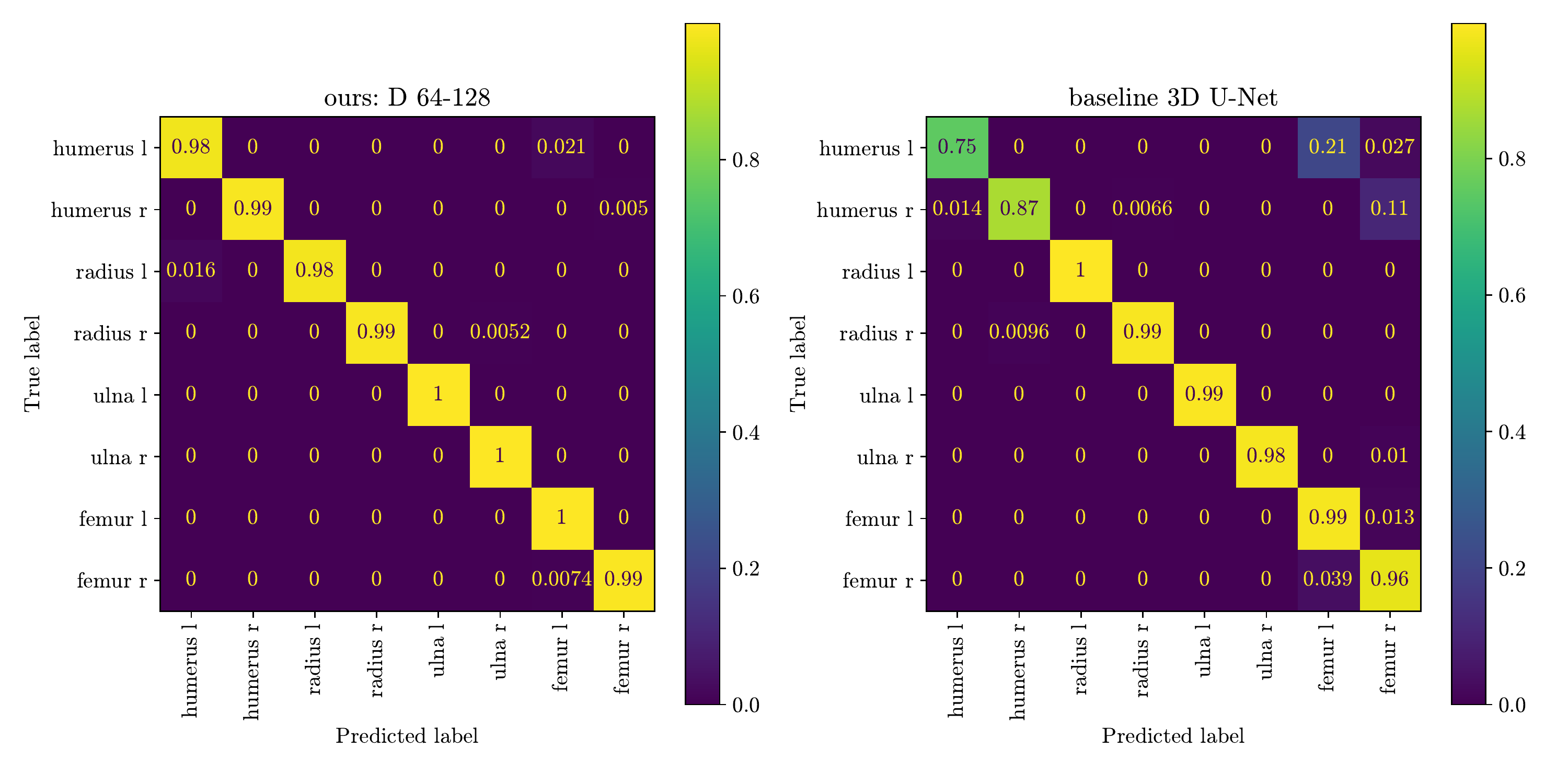}
\caption{Confusion matrix among the long bones of the arms and legs. With our method, there is considerably less confusion between the left and right sides of the body and between arm and leg bones.}
\label{fig_confusion_matrix}
\end{figure*}

Our experiments show how automated distinct bone segmentation can be improved using a multi-resolution approach. We evaluate our results on multiple target resolutions with different numbers of context networks and field of view sizes and perform an ablation study to determine the most beneficial way to combine context and target network information.

We evaluated some of the most common errors when using a baseline segmentation method. We found that the missing context information leads to similar-looking bones in different body regions being mistaken for one another. In the confusion matrix presented in \autoref{fig_confusion_matrix}, we observe that when using a baseline 3D U-Net, humerus pixels were predicted as femur, and the left and right humerus were confused for one another (right confusion matrix). When using context information, these errors are reduced almost entirely (left confusion matrix).

\begin{table}[!t]
\caption{\label{tab_ablation}Ablation results in DSC for different model configurations.}
\centering
\begin{tabular}{llrrrclrrr}
\toprule
\multicolumn{1}{l}{Target patch size}    & \multicolumn{4}{c}{32} && \multicolumn{4}{c}{64} \\
\cmidrule{2-5} \cmidrule{7-10}
  &  &   &  & non-zero  &&  & &   & non-zero\\
DSC & Median & $\sigma$ & $-\sigma$ & DSC && Median & $\sigma$ & $-\sigma$ &  DSC\\
\midrule
A 3D U-Net & 0.64&+0.19&-0.34&94.5\% && 0.83&+0.09&-0.27&94.5\% \\ 
B HookNet & 0.66&+0.17&-0.34&94.1\%  && 0.85&+0.09&-0.32&95.3\% \\ 
C MRN & 0.69&+0.16&-0.37&95.1\% && 0.84&+0.09&-0.31&96.0\% \\ 
D SneakyNet (ours) & 0.75&+0.14&-0.33&95.3\% && 0.86&+0.08&-0.28&96.7\% \\ 
\bottomrule
\end{tabular}
\end{table}

We performed an ablation study to see how different strategies of combining the context and target information within the network perform. In \autoref{tab_ablation} we present the quantitative results.
For both target patch sizes, 32 and 64, all strategies (B-D) improve upon the baseline 3D U-Net (A). 
The observed effect is substantially bigger when using the smaller target patch size of $32^3$, where the median DCS rises from 0.64 to 0.75. The DSC still increases from 0.83 to 0.86 median DSC on the bigger target patches.

The combination of skip connections at every level and a context loss function in our proposed architecture increases the accuracy further, as compared to the HookNet \cite{van2021hooknet} and the MRN \cite{gu2018multi}. 

\begin{figure*}[!t]
\centering
\includegraphics[width=\textwidth]{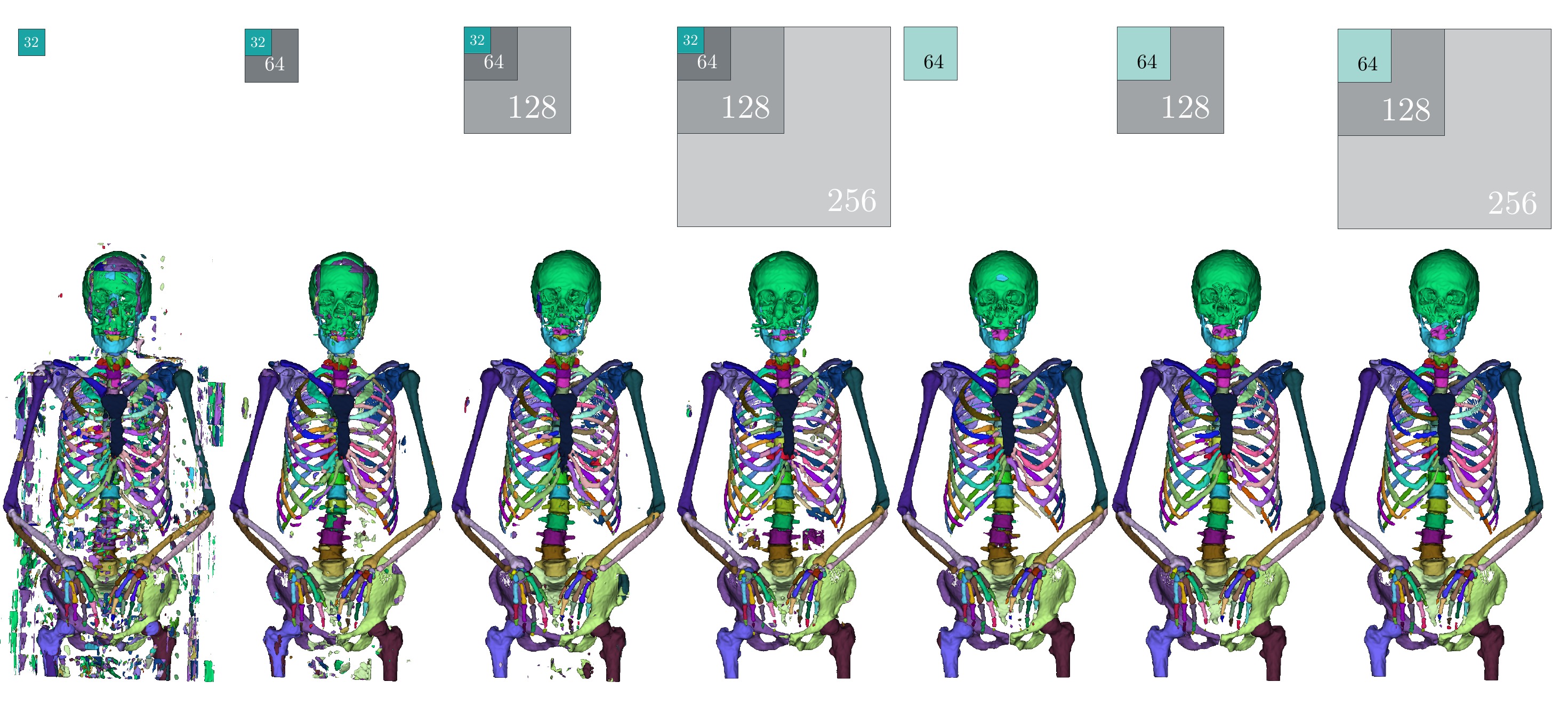}
\caption{Qualitative prediction results from our ablation study comparing different numbers of context networks at various resolutions. The first four results from the left were obtained using a target patch size of 32px per dimension (turquoise), and the remaining three scans with target patch sizes of 64px per dimension (light blue). The grey areas indicate the field of view of the context networks. The sizes of the squares are proportional to the prediction sizes.}
\label{fig_seg_size}
\end{figure*}

In \autoref{tab_context_branch_comparison} 
we ablate the influence of different numbers of context networks and input patch sizes. Qualitative results are depicted in \autoref{fig_seg_size}. 
Comparing the baseline 3D U-Nets with the SneakyNet results, we see that adding context networks to very small target patches of $32^3$ pixels almost reaches the performance of our baseline networks operating on $64^3$ patches. Going up, the SneakyNet operating on patch size $64^3$ even outperforms the baseline 3D U-Net of patchsize $128^3$. We recall, that we had to reduce the number of channels in the baseline $128^3$ network, due to memory restraints. Our ablation results suggests, that the addition of context networks are more valuable in adding performance when reaching memory limits.
When considering the different FOV of the context networks, we observe the best results when including context FOVs of $128^3$. This covers roughly half of the L-R and A-P dimensions of the scans and seems to contain the necessary information to correctly locate bones, see e.g. the purple lumbar vertebra in \autoref{fig_seg_size}, which is correctly located in cases where the context FOV reaches $128^3$.

We provide a comparison to other results published on distinct bone segmentation in \autoref{tab_comparison_others}. While a direct comparison is difficult due to different datasets, our results compare favourably to both the convolutional neural networks and shape model approach by \cite{lindgren_belal_deep_2019}, and to the hierarchical atlas segmentation by \cite{fu_hierarchical_2017}.

\begin{table}[!t]
\caption{\label{tab_context_branch_comparison}Ablation results for the number of context networks in the SneakyNet architecture (D). Zero context networks corresponds to the baseline 3D U-Nets (A) with different input patch sizes.}
\centering
\begin{tabular}{llllrrr}
\toprule
Config& target FOV&context FOV(s)  & \multicolumn{4}{c}{DSC} \\
\cmidrule{4-7}
&per dim.&per dim. & Median & $\sigma$ & $-\sigma$ & non-zero DSC\\
\midrule

A&32& --- & 0.64&+0.19&-0.34&94.5\% \\
D&32&64 & 0.75&+0.14&-0.33&95.3\% \\
D&32&64-128 & $\mathbf{0.79}$&+0.11&-0.33&94.4\% \\
D&32&64-128-256 & 0.79&+0.11&-0.33&95.9\% \\

\midrule
A&64& ---& 0.83&+0.09&-0.27&95.6\% \\
D&64&128 & $\mathbf{0.86}$&+0.08&-0.28&96.7\% \\
D&64&128-256 & 0.85&+0.09&-0.28&96.1\% \\
\midrule
A&128&--- & 0.82&+0.11&-0.30&94.3\% \\
\bottomrule
\end{tabular}
\end{table}

\begin{table}[t]
\caption{Comparison of our best-performing SneakyNet (D, target patch size of $64^3$ and one context network with a FOV of $128^3$ pixels) to other work on distinct bone segmentation from upper-body CT. Results are in DSC.}\label{tab_comparison_others}
\begin{tabular}{l r r r }
\toprule
\multicolumn{1}{l}{}& \multicolumn{1}{c}{ours (median)} & \multicolumn{1}{c}{\cite{lindgren_belal_deep_2019} (median)} & \multicolumn{1}{c}{\cite{fu_hierarchical_2017} (mean)} \\
\midrule
L3   & 0.85  & 0.85 & 0.91 \\

Sacrum   & 0.92 & 0.88 &    \\

Right $7^{\textrm{th}}$ rib &0.78 &0.84 & \\

Clavicula & 0.94 &  & 0.87 \\

Right femur & 0.96 &  & 0.92  \\
Pelvic bones & 0.94 &  & 0.86  \\

Hamate & 0.81 &  &   \\

\midrule
Inference time per scan (min)& \multicolumn{1}{r}{$\sim3$} &
\multicolumn{1}{r}{ } & \multicolumn{1}{r}{$\sim 20$} \\
Scans in training dataset (\#) & \multicolumn{1}{r}{11} &
\multicolumn{1}{r}{100} & \multicolumn{1}{r}{19} \\
Classes (\#)& \multicolumn{1}{r}{125} & \multicolumn{1}{r}{49} & \multicolumn{1}{r}{62} \\
\bottomrule
\end{tabular}
\end{table}


\section{Conclusion}
This works presents improvements in distinct bone segmentation from upper-body CT. The proposed multi-resolution networks use additional inputs at a lower resolution but with a larger field of view to provide the necessary context information to assign the proper bone classes. We compared three different ways of combining the context and target information and evaluated the results using zero to three context networks. Using context networks improves the segmentation results on all target patch sizes.


\backmatter





\bmhead{Acknowledgments}
This work was financially supported by the Werner Siemens Foundation through the MIRACLE project. We thank Azhar Zam for valuable discussions that helped shape this work.



\section*{Declarations}

\begin{itemize}
\item Funding:
This work was financially supported by the Werner Siemens
Foundation through the MIRACLE project.
\item Competing interests:
None of the authors have competing interests to declare that are relevant to the content of this article.
\item Ethics approval:
This research study was conducted retrospectively from CT data routinely obtained from body donours. No ethical approval is required.
\item Consent to participate:
Informed consent was obtained from all individual body donours included in the study.
\item Consent for publication:
Body donours signed informed consent regarding publications using their data.
\item Availability of data and materials:
The upper-body CT dataset is not publicly available. An anonymized version can be shared on request.
\item Code availability:
Our code is shared at: \url{https://gitlab.com/cian.unibas.ch/sneakynet}
\end{itemize}

\bibliography{sn-bibliography}


\begin{thebibliography}{21}
\ifx \bisbn   \undefined \def \bisbn  #1{ISBN #1}\fi
\ifx \binits  \undefined \def \binits#1{#1}\fi
\ifx \bauthor  \undefined \def \bauthor#1{#1}\fi
\ifx \batitle  \undefined \def \batitle#1{#1}\fi
\ifx \bjtitle  \undefined \def \bjtitle#1{#1}\fi
\ifx \bvolume  \undefined \def \bvolume#1{\textbf{#1}}\fi
\ifx \byear  \undefined \def \byear#1{#1}\fi
\ifx \bissue  \undefined \def \bissue#1{#1}\fi
\ifx \bfpage  \undefined \def \bfpage#1{#1}\fi
\ifx \blpage  \undefined \def \blpage #1{#1}\fi
\ifx \burl  \undefined \def \burl#1{\textsf{#1}}\fi
\ifx \doiurl  \undefined \def \doiurl#1{\url{https://doi.org/#1}}\fi
\ifx \betal  \undefined \def \betal{\textit{et al.}}\fi
\ifx \binstitute  \undefined \def \binstitute#1{#1}\fi
\ifx \binstitutionaled  \undefined \def \binstitutionaled#1{#1}\fi
\ifx \bctitle  \undefined \def \bctitle#1{#1}\fi
\ifx \beditor  \undefined \def \beditor#1{#1}\fi
\ifx \bpublisher  \undefined \def \bpublisher#1{#1}\fi
\ifx \bbtitle  \undefined \def \bbtitle#1{#1}\fi
\ifx \bedition  \undefined \def \bedition#1{#1}\fi
\ifx \bseriesno  \undefined \def \bseriesno#1{#1}\fi
\ifx \blocation  \undefined \def \blocation#1{#1}\fi
\ifx \bsertitle  \undefined \def \bsertitle#1{#1}\fi
\ifx \bsnm \undefined \def \bsnm#1{#1}\fi
\ifx \bsuffix \undefined \def \bsuffix#1{#1}\fi
\ifx \bparticle \undefined \def \bparticle#1{#1}\fi
\ifx \barticle \undefined \def \barticle#1{#1}\fi
\bibcommenthead
\ifx \bconfdate \undefined \def \bconfdate #1{#1}\fi
\ifx \botherref \undefined \def \botherref #1{#1}\fi
\ifx \url \undefined \def \url#1{\textsf{#1}}\fi
\ifx \bchapter \undefined \def \bchapter#1{#1}\fi
\ifx \bbook \undefined \def \bbook#1{#1}\fi
\ifx \bcomment \undefined \def \bcomment#1{#1}\fi
\ifx \oauthor \undefined \def \oauthor#1{#1}\fi
\ifx \citeauthoryear \undefined \def \citeauthoryear#1{#1}\fi
\ifx \endbibitem  \undefined \def \endbibitem {}\fi
\ifx \bconflocation  \undefined \def \bconflocation#1{#1}\fi
\ifx \arxivurl  \undefined \def \arxivurl#1{\textsf{#1}}\fi
\csname PreBibitemsHook\endcsname

\bibitem{deng2022deep}
\begin{botherref}
\oauthor{\bsnm{Deng}, \binits{Y.}},
\oauthor{\bsnm{Wang}, \binits{L.}},
\oauthor{\bsnm{Zhao}, \binits{C.}},
\oauthor{\bsnm{Tang}, \binits{S.}},
\oauthor{\bsnm{Cheng}, \binits{X.}},
\oauthor{\bsnm{Deng}, \binits{H.-W.}},
\oauthor{\bsnm{Zhou}, \binits{W.}}:
A deep learning-based approach to automatic proximal femur segmentation in
  quantitative ct images.
Medical \& Biological Engineering \& Computing,
1--13
(2022)
\end{botherref}
\endbibitem

\bibitem{uemura2022development}
\begin{barticle}
\bauthor{\bsnm{Uemura}, \binits{K.}},
\bauthor{\bsnm{Otake}, \binits{Y.}},
\bauthor{\bsnm{Takao}, \binits{M.}},
\bauthor{\bsnm{Makino}, \binits{H.}},
\bauthor{\bsnm{Soufi}, \binits{M.}},
\bauthor{\bsnm{Iwasa}, \binits{M.}},
\bauthor{\bsnm{Sugano}, \binits{N.}},
\bauthor{\bsnm{Sato}, \binits{Y.}}:
\batitle{Development of an open-source measurement system to assess the areal
  bone mineral density of the proximal femur from clinical ct images}.
\bjtitle{Archives of Osteoporosis}
\bvolume{17}(\bissue{1}),
\bfpage{1}--\blpage{11}
(\byear{2022})
\end{barticle}
\endbibitem

\bibitem{su2022three}
\begin{botherref}
\oauthor{\bsnm{Su}, \binits{Z.}},
\oauthor{\bsnm{Liu}, \binits{Z.}},
\oauthor{\bsnm{Wang}, \binits{M.}},
\oauthor{\bsnm{Li}, \binits{S.}},
\oauthor{\bsnm{Lin}, \binits{L.}},
\oauthor{\bsnm{Yuan}, \binits{Z.}},
\oauthor{\bsnm{Pang}, \binits{S.}},
\oauthor{\bsnm{Feng}, \binits{Q.}},
\oauthor{\bsnm{Chen}, \binits{T.}},
\oauthor{\bsnm{Lu}, \binits{H.}}:
Three-dimensional reconstruction of kambin's triangle based on automated
  magnetic resonance image segmentation.
Journal of Orthopaedic Research{\textregistered}
(2022)
\end{botherref}
\endbibitem

\bibitem{klein2019automatic}
\begin{barticle}
\bauthor{\bsnm{Klein}, \binits{A.}},
\bauthor{\bsnm{Warszawski}, \binits{J.}},
\bauthor{\bsnm{Hillenga{\ss}}, \binits{J.}},
\bauthor{\bsnm{Maier-Hein}, \binits{K.H.}}:
\batitle{Automatic bone segmentation in whole-body ct images}.
\bjtitle{International journal of computer assisted radiology and surgery}
\bvolume{14}(\bissue{1}),
\bfpage{21}--\blpage{29}
(\byear{2019})
\end{barticle}
\endbibitem

\bibitem{leydon2021bone}
\begin{botherref}
\oauthor{\bsnm{Leydon}, \binits{P.}},
\oauthor{\bsnm{O'Connell}, \binits{M.}},
\oauthor{\bsnm{Greene}, \binits{D.}},
\oauthor{\bsnm{Curran}, \binits{K.M.}}:
Bone segmentation in contrast enhanced whole-body computed tomography.
Biomedical Physics \& Engineering Express
(2021)
\end{botherref}
\endbibitem

\bibitem{noguchi2020bone}
\begin{barticle}
\bauthor{\bsnm{Noguchi}, \binits{S.}},
\bauthor{\bsnm{Nishio}, \binits{M.}},
\bauthor{\bsnm{Yakami}, \binits{M.}},
\bauthor{\bsnm{Nakagomi}, \binits{K.}},
\bauthor{\bsnm{Togashi}, \binits{K.}}:
\batitle{Bone segmentation on whole-body ct using convolutional neural network
  with novel data augmentation techniques}.
\bjtitle{Computers in biology and medicine}
\bvolume{121},
\bfpage{103767}
(\byear{2020})
\end{barticle}
\endbibitem

\bibitem{payer2020coarse}
\begin{bchapter}
\bauthor{\bsnm{Payer}, \binits{C.}},
\bauthor{\bsnm{Stern}, \binits{D.}},
\bauthor{\bsnm{Bischof}, \binits{H.}},
\bauthor{\bsnm{Urschler}, \binits{M.}}:
\bctitle{Coarse to fine vertebrae localization and segmentation with
  spatialconfiguration-net and u-net.}
In: \bbtitle{VISIGRAPP (5: VISAPP)},
pp. \bfpage{124}--\blpage{133}
(\byear{2020})
\end{bchapter}
\endbibitem

\bibitem{yang2021ribseg}
\begin{bchapter}
\bauthor{\bsnm{Yang}, \binits{J.}},
\bauthor{\bsnm{Gu}, \binits{S.}},
\bauthor{\bsnm{Wei}, \binits{D.}},
\bauthor{\bsnm{Pfister}, \binits{H.}},
\bauthor{\bsnm{Ni}, \binits{B.}}:
\bctitle{Ribseg dataset and strong point cloud baselines for rib segmentation
  from ct scans}.
In: \bbtitle{International Conference on Medical Image Computing and
  Computer-Assisted Intervention},
pp. \bfpage{611}--\blpage{621}
(\byear{2021}).
\bcomment{Springer}
\end{bchapter}
\endbibitem

\bibitem{faisal2021carpal}
\begin{botherref}
\oauthor{\bsnm{Faisal}, \binits{A.}},
\oauthor{\bsnm{Khalil}, \binits{A.}},
\oauthor{\bsnm{Chai}, \binits{H.Y.}},
\oauthor{\bsnm{Lai}, \binits{K.W.}}:
X-ray carpal bone segmentation and area measurement.
Multimedia Tools and Applications,
1--12
(2021)
\end{botherref}
\endbibitem

\bibitem{fu_hierarchical_2017}
\begin{barticle}
\bauthor{\bsnm{Fu}, \binits{Y.}},
\bauthor{\bsnm{Liu}, \binits{S.}},
\bauthor{\bsnm{Li}, \binits{H.H.}},
\bauthor{\bsnm{Yang}, \binits{D.}}:
\batitle{Automatic and hierarchical segmentation of the human skeleton in {CT}
  images}.
\bjtitle{Physics in Medicine and Biology}
\bvolume{62}(\bissue{7}),
\bfpage{2812}--\blpage{2833}
(\byear{2017})
\end{barticle}
\endbibitem

\bibitem{lindgren_belal_deep_2019}
\begin{barticle}
\bauthor{\bsnm{Lindgren~Belal}, \binits{S.}},
\bauthor{\bsnm{Sadik}, \binits{M.}},
\bauthor{\bsnm{Kaboteh}, \binits{R.}},
\bauthor{\bsnm{Enqvist}, \binits{O.}},
\bauthor{\bsnm{Ulén}, \binits{J.}},
\bauthor{\bsnm{Poulsen}, \binits{M.H.}},
\bauthor{\bsnm{Simonsen}, \binits{J.}},
\bauthor{\bsnm{Høilund-Carlsen}, \binits{P.F.}},
\bauthor{\bsnm{Edenbrandt}, \binits{L.}},
\bauthor{\bsnm{Trägårdh}, \binits{E.}}:
\batitle{Deep learning for segmentation of 49 selected bones in {CT} scans:
  {First} step in automated {PET}/{CT}-based {3D} quantification of skeletal
  metastases}.
\bjtitle{European Journal of Radiology}
\bvolume{113},
\bfpage{89}--\blpage{95}
(\byear{2019})
\end{barticle}
\endbibitem

\bibitem{kamiya2018automated}
\begin{bchapter}
\bauthor{\bsnm{Kamiya}, \binits{N.}},
\bauthor{\bsnm{Kume}, \binits{M.}},
\bauthor{\bsnm{Zheng}, \binits{G.}},
\bauthor{\bsnm{Zhou}, \binits{X.}},
\bauthor{\bsnm{Kato}, \binits{H.}},
\bauthor{\bsnm{Chen}, \binits{H.}},
\bauthor{\bsnm{Muramatsu}, \binits{C.}},
\bauthor{\bsnm{Hara}, \binits{T.}},
\bauthor{\bsnm{Miyoshi}, \binits{T.}},
\bauthor{\bsnm{Matsuo}, \binits{M.}},
\bauthor{\bsnm{Fujita}, \binits{H.}}:
\bctitle{Automated recognition of erector spinae muscles and their skeletal
  attachment region via deep learning in torso ct images}.
In: \bbtitle{International Workshop on Computational Methods and Clinical
  Applications in Musculoskeletal Imaging},
pp. \bfpage{1}--\blpage{10}
(\byear{2018}).
\bcomment{Springer}
\end{bchapter}
\endbibitem

\bibitem{boutillon2022multi}
\begin{barticle}
\bauthor{\bsnm{Boutillon}, \binits{A.}},
\bauthor{\bsnm{Borotikar}, \binits{B.}},
\bauthor{\bsnm{Burdin}, \binits{V.}},
\bauthor{\bsnm{Conze}, \binits{P.-H.}}:
\batitle{Multi-structure bone segmentation in pediatric mr images with combined
  regularization from shape priors and adversarial network}.
\bjtitle{Artificial Intelligence in Medicine}
\bvolume{132},
\bfpage{102364}
(\byear{2022})
\end{barticle}
\endbibitem

\bibitem{isensee2021nnu}
\begin{barticle}
\bauthor{\bsnm{Isensee}, \binits{F.}},
\bauthor{\bsnm{Jaeger}, \binits{P.F.}},
\bauthor{\bsnm{Kohl}, \binits{S.A.}},
\bauthor{\bsnm{Petersen}, \binits{J.}},
\bauthor{\bsnm{Maier-Hein}, \binits{K.H.}}:
\batitle{nnu-net: a self-configuring method for deep learning-based biomedical
  image segmentation}.
\bjtitle{Nature Methods}
\bvolume{18}(\bissue{2}),
\bfpage{203}--\blpage{211}
(\byear{2021})
\end{barticle}
\endbibitem

\bibitem{gu2018multi}
\begin{botherref}
\oauthor{\bsnm{Gu}, \binits{F.}},
\oauthor{\bsnm{Burlutskiy}, \binits{N.}},
\oauthor{\bsnm{Andersson}, \binits{M.}},
\oauthor{\bsnm{Wil{\'e}n}, \binits{L.K.}}:
Multi-resolution networks for semantic segmentation in whole slide images.
Computational Pathology and Ophthalmic Medical Image Analysis,
11--18
(2018)
\end{botherref}
\endbibitem

\bibitem{van2021hooknet}
\begin{barticle}
\bauthor{\bsnm{Van~Rijthoven}, \binits{M.}},
\bauthor{\bsnm{Balkenhol}, \binits{M.}},
\bauthor{\bsnm{Sili{\c{n}}a}, \binits{K.}},
\bauthor{\bsnm{Van Der~Laak}, \binits{J.}},
\bauthor{\bsnm{Ciompi}, \binits{F.}}:
\batitle{Hooknet: Multi-resolution convolutional neural networks for semantic
  segmentation in histopathology whole-slide images}.
\bjtitle{Medical Image Analysis}
\bvolume{68},
\bfpage{101890}
(\byear{2021})
\end{barticle}
\endbibitem

\bibitem{faludi2021transfer}
\begin{botherref}
\oauthor{\bsnm{Faludi}, \binits{B.}},
\oauthor{\bsnm{Zentai}, \binits{N.}},
\oauthor{\bsnm{Zelechowski}, \binits{M.}},
\oauthor{\bsnm{Zam}, \binits{A.}},
\oauthor{\bsnm{Rauter}, \binits{G.}},
\oauthor{\bsnm{Griessen}, \binits{M.}},
\oauthor{\bsnm{Cattin}, \binits{P.C.}}:
Transfer-function-independent acceleration structure for volume rendering in
  virtual reality
(2021)
\end{botherref}
\endbibitem

\bibitem{zelechowski2021patient}
\begin{botherref}
\oauthor{\bsnm{{\.Z}elechowski}, \binits{M.}},
\oauthor{\bsnm{Karnam}, \binits{M.}},
\oauthor{\bsnm{Faludi}, \binits{B.}},
\oauthor{\bsnm{Gerig}, \binits{N.}},
\oauthor{\bsnm{Rauter}, \binits{G.}},
\oauthor{\bsnm{Cattin}, \binits{P.C.}}:
Patient positioning by visualising surgical robot rotational workspace in
  augmented reality.
Computer Methods in Biomechanics and Biomedical Engineering: Imaging \&
  Visualization,
1--7
(2021)
\end{botherref}
\endbibitem

\bibitem{schnider2022improved}
\begin{botherref}
\oauthor{\bsnm{Schnider}, \binits{E.}},
\oauthor{\bsnm{Huck}, \binits{A.}},
\oauthor{\bsnm{Toranelli}, \binits{M.}},
\oauthor{\bsnm{Rauter}, \binits{G.}},
\oauthor{\bsnm{M{\"u}ller-Gerbl}, \binits{M.}},
\oauthor{\bsnm{Cattin}, \binits{P.C.}}:
Improved distinct bone segmentation from upper-body ct using
  binary-prediction-enhanced multi-class inference.
International Journal of Computer Assisted Radiology and Surgery,
1--8
(2022)
\end{botherref}
\endbibitem

\bibitem{schnider20203d}
\begin{bchapter}
\bauthor{\bsnm{Schnider}, \binits{E.}},
\bauthor{\bsnm{Horv{\'a}th}, \binits{A.}},
\bauthor{\bsnm{Rauter}, \binits{G.}},
\bauthor{\bsnm{Zam}, \binits{A.}},
\bauthor{\bsnm{M{\"u}ller-Gerbl}, \binits{M.}},
\bauthor{\bsnm{Cattin}, \binits{P.C.}}:
\bctitle{3d segmentation networks for excessive numbers of classes: Distinct
  bone segmentation in upper bodies}.
In: \bbtitle{International Workshop on Machine Learning in Medical Imaging},
pp. \bfpage{40}--\blpage{49}
(\byear{2020}).
\bcomment{Springer}
\end{bchapter}
\endbibitem

\bibitem{milletari2016v}
\begin{bchapter}
\bauthor{\bsnm{Milletari}, \binits{F.}},
\bauthor{\bsnm{Navab}, \binits{N.}},
\bauthor{\bsnm{Ahmadi}, \binits{S.-A.}}:
\bctitle{V-net: Fully convolutional neural networks for volumetric medical
  image segmentation}.
In: \bbtitle{2016 Fourth International Conference on 3D Vision (3DV)},
pp. \bfpage{565}--\blpage{571}
(\byear{2016}).
\bcomment{IEEE}
\end{bchapter}
\endbibitem

\end{thebibliography}


\end{document}